\documentclass[lettersize,journal]{IEEEtran}
\usepackage{amsmath,amsfonts}
\usepackage{algorithmic}
\usepackage{algorithm}
\usepackage{array}
\usepackage{subfigure} 
\usepackage{svg}
\usepackage{textcomp}
\usepackage{stfloats}
\usepackage{url}
\usepackage{verbatim}
\usepackage{xcolor}
\usepackage{graphicx}
\usepackage{multirow}
\usepackage{array}
\usepackage{threeparttable}
\usepackage{tabularx}
\usepackage{booktabs}
\usepackage{makecell}
\newcolumntype{P}[1]{>{\centering\arraybackslash}p{#1}}
\newcolumntype{M}[1]{>{\centering\arraybackslash}m{#1}}

\usepackage{cite}
\usepackage{lipsum}

\begin{document}

\title{Communication Optimization for Distributed Training: Architecture, Advances, and Opportunities}




\author{Yunze Wei, Tianshuo Hu, Cong Liang, Yong Cui,~\IEEEmembership{Member,~IEEE}
  \IEEEcompsocitemizethanks
  {
   \IEEEcompsocthanksitem Y.~Wei, T.~Hu, C.~Liang and Y.~Cui are with the Department of Computer Science and Technology, Tsinghua University, Beijing, China.
  \IEEEcompsocthanksitem Yong Cui~(cuiyong@tsinghua.edu.cn) is the corresponding author.
  }
}

\maketitle

\begin{abstract}

The past few years have witnessed the flourishing of large-scale deep neural network models with ever-growing parameter numbers.
Training such large-scale models typically requires massive memory and computing resources, necessitating distributed training.
As GPU performance has rapidly evolved in recent years, computation time has shrunk, 
making communication a larger portion of the overall training time.
Consequently, optimizing communication for distributed training has become crucial.
In this article, we briefly introduce the general architecture of distributed deep neural network training and analyze relationships among \textit{Parallelization Strategy, Collective Communication Library, and Network} from the perspective of communication optimization, which forms a three-layer paradigm.
We then review current representative research advances within this three-layer paradigm.
We find that layers in the current three-layer paradigm are relatively independent and there is a rich design space for cross-layer collaborative optimization in distributed training scenarios.
Therefore, we advocate \textit{``Vertical" and ``Horizontal"} co-designs which extend the three-layer paradigm to a five-layer paradigm.
We also advocate \textit{``Intra-Inter" and ``Host-Net"} co-designs to further utilize the potential of heterogeneous resources.
We hope this article can shed some light on future research on communication optimization for distributed training.


\end{abstract}


\section{Introduction}\label{sec:intro}

\IEEEPARstart{L}{arge-scale} deep neural network (DNN) models (``large models" for short) have become ubiquitous, and their capabilities have significantly advanced in recent years.
The latest prominent large models such as GPT\footnote{https://openai.com/gpt-4}, LLaMA\footnote{https://llama.meta.com}, and GLM\footnote{https://open.bigmodel.cn} demonstrated unprecedented performance, enabling significant changes in production and lifestyle.

The rapid development of large models leads to a significant increase in the scale of parameters and training data.
Due to the limited memory and computing power of a single GPU, training large models with a single GPU has already been a thing of the past. 
For instance, training a GPT-3 model with 175 billion parameters on a single Nvidia V100 GPU would require approximately 288 years~\cite{narayanan2021efficient}. 
Consequently, training large models with multiple GPUs is natural and inevitable.
This creates a new demand for large-scale, high-performance GPU clusters to accelerate the model training process. 

However, simply enhancing GPU performance and enlarging cluster size do not necessarily lead to linear performance improvement of distributed deep neural network training (``distributed training" for short) systems, as one might naturally expect.
This is because the time overhead of distributed training comes not only from computation but also from communication.
When the time spent on computation is reduced, the communication time is exposed more and gradually becomes a bottleneck. 
The communication overhead can account for up to 60\% of a DNN training iteration time in Meta’s production environment~\cite{wang2023topoopt}. 
Therefore, there is an urgent need to enhance communication efficiency for distributed training.

Many components of distributed training systems\cite{jiang2020unified, jouppi2023tpu} are closely related to communication. 
For instance, parallelization strategy determines communication demand, collective communication library (CCL) generates communication traffic, and network affects the efficiency of performing communication tasks.
To sum up, there is a \textit{Parallelization Strategy, CCL, and Network} three-layer paradigm of communication optimization for distributed training. 

There are many research advances optimizing communication for distributed training in the current three-layer paradigm.
For example, PTD-P~\cite{narayanan2021efficient} uses a novel interleaved pipeline parallelism optimization scheme to overlap communication and computation as much as possible.
TACCL~\cite{shah2023taccl} generates communication primitive algorithms tailored to specific training tasks and topology to enhance efficiency. 
TopoOpt~\cite{wang2023topoopt} leverages the reconfigurability of optical switches to optimize topologies and parallel strategies collaboratively, providing customized topology for efficient communication. 

Unlike general high-performance computing scenarios, distributed training has distinct characteristics. 
For instance, the communication traffic of each iteration is typically fixed~\cite{rajasekaran2024cassini} and most of the communication tasks are pre-determined, which contrasts with the stochastic traffic in traditional scenarios. 
However, components in the current communication paradigm often function independently, hindering collaborative optimization for communication. 
In Section~\ref{sec:oppo}, we explore research opportunities from a collaborative design perspective to reduce job completion time (JCT) and enhance the efficiency of distributed training systems.

The rest of this article is arranged as follows: Section~\ref{sec:arc} presents the general architecture of distributed training and the current three-layer paradigm from the perspective of communication optimization. 
Section~\ref{sec:adv} reviews representative research advances.
Section~\ref{sec:oppo} illustrates the necessity of collaborative design through a case study and provides an outlook on research opportunities.
Finally is the conclusion.

\section{Architecture}\label{sec:arc}

\begin{figure}[t]
    \centering
    \includegraphics[width=0.75\linewidth]{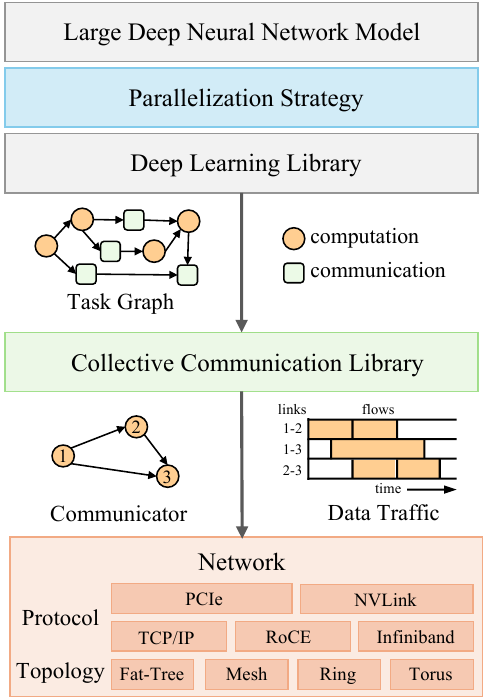}
    \caption{Distributed training architecture from a communication perspective.}
    \label{fig:arc-new}
\end{figure}

This section first provides an overview of the distributed training architecture from the perspective of communication, followed by an in-depth introduction of three components pertinent to communication optimization: parallelization strategy, CCL, and network. 
Finally, we summarize the three components as a three-layer paradigm of communication optimization and analyze their impacts on communication efficiency.

\subsection{Overview}
From the perspective of communication, the general architecture of the distributed training system is shown in Fig.~\ref{fig:arc-new}.
Top of the architecture is a \textit{large deep neural network model} that requires distributed training.

\textit{Parallelization strategy} is an essential part of distributed training deployment which determines how a model is partitioned for distributed training. Commonly used parallelization strategies are detailed in Section~\ref{subsec:cp}. 

Deep learning models are often implemented with the \textit{deep learning library} (such as TensorFlow\footnote{https://www.tensorflow.org} and PyTorch\footnote{https://pytorch.org})
, which generates execution \textit{task graphs} including computing tasks and corresponding collective communication tasks.

The deep learning library often invokes \textit{collective communication library} to implement collective communication primitives to transmit activations (in forward propagation) or synchronize gradients (in backward propagation) between different GPUs. 
CCL often gathers several GPU nodes as a communicator and schedules data traffic to finish a collective communication task.
Commonly used collective communication primitives are detailed in Section~\ref{subsec:ccp}.

The actual communication \textit{data traffic} generated by CCL is injected into the underlying \textit{network}. A variety of network protocols and topologies for distributed training are detailed in Section~\ref{subsec:network}. 

Parallelization strategy, CCL, and network are three critical components in the architecture that affect communication efficiency. The interplay among them and their impacts on communication efficiency are analyzed in Section~\ref{subsec:comm_in_arc}.

In order to reduce communication overhead, various communication compression methods are also commonly used in practice but often sacrifice accuracy or memory. 
Communication compression methods are often associated with deep learning libraries and are out of the scope of this article due to space constraints.

\begin{figure*}[t]
  \centering
  \begin{minipage}[b]{0.34\textwidth}
  \centering
      \begin{minipage}[b]{\textwidth}
        \includegraphics[width=0.95\textwidth]{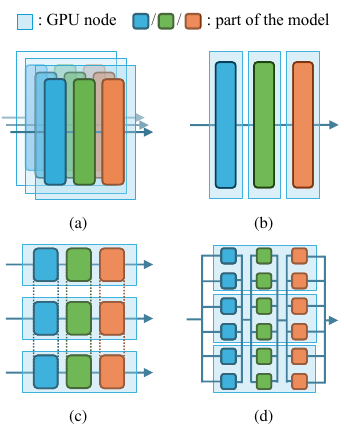}
        \caption{Four common parallelization strategies: a) Data parallelism; b) Pipeline parallelism; c) Tensor parallelism; d) MoE parallelism. 
        }
        \label{fig:para}
      \end{minipage}
    \end{minipage}
    \hspace{5pt}
    \begin{minipage}[b]{0.63\textwidth}
      \begin{minipage}[b]{\textwidth}
        \includegraphics[width=0.95\textwidth]{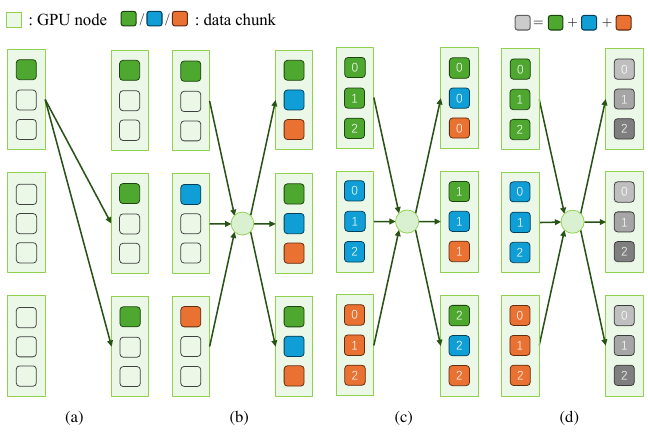}
        \caption{Four common collective communication primitives: a) BroadCast; b) All-Gather; c) All-to-All; d) All-Reduce. The left side of each figure represents the data state before communication, and the right side represents the state after communication. 
        }
        \label{fig:ccl}
      \end{minipage}
  \end{minipage}
\end{figure*}

\subsection{Common Parallelization Strategies}\label{subsec:cp}
Fig. 2 shows four commonly used parallelization strategies that include data parallelism, 
model parallelism (includes pipeline and tensor parallelism), and emerging Mixture-of-Expert (MoE) parallelism. 

\textit{Data parallelism} is one of the most commonly used parallelization strategies that distribute a subset of training data (mini-batch) to multiple copies of a model on different GPUs.  
\textit{Model parallelism}, including pipeline and tensor parallelism, means splitting the model onto different GPUs. 
\textit{Pipeline parallelism} allocates different layers of the model to different GPUs, so there is mainly point-to-point communication between layers (GPUs). 
\textit{Tensor parallelism} splits the same layer of the model onto different GPUs and uses distributed matrix computing techniques for collaboration~\cite{shoeybi2019megatron}, which is a communication-intensive operation.
\textit{MoE parallelism} involves dividing a portion of a model into multiple expert components. 
Each expert specializes in a specific task domain and is allocated to a certain GPU.

In practice, the above parallelization strategies are often not used independently, but in a hybrid manner. 
For example, we can first divide the model by layers and split each layer, then distribute the divided model to multiple sets of GPUs, thereby achieving pipeline-tensor-data three-dimensional hybrid parallelism~\cite{narayanan2021efficient}. 
MoE parallelism is another example of hybrid parallelism. The idea of data parallelism is implicit in MoE parallelism, and the expert model itself can also be split onto multiple GPUs for parallel computation.

\subsection{Common Collective Communication Primitives}\label{subsec:ccp}
Commonly used collective communication primitives implemented in CCL are illustrated in Fig.~\ref{fig:ccl}.

\textit{Broadcast} distributes data from a particular node to all other nodes, which can be used in data parallelism and some model parallelism~\cite{zhuang2023optimizing}.
\textit{All-Gather} is a many-to-many collective communication primitive where data from different nodes are distributed to all nodes.
\textit{All-to-All} transmits data among various nodes, such as data distribution in MoE parallelism~\cite{li2023accelerating, liu2023janus}.
\textit{All-Reduce} is a sum operation of the corresponding data chunk of each node, as shown in Fig.~\ref{fig:ccl}(d), where each grey rounded rectangle represents the aggregated results of the data chunks at the corresponding position of each GPU node. Common All-Reduce scenarios include data parallelism and some model parallelism such as Megatron-lm~\cite{shoeybi2019megatron}.

\subsection{Underlying Network}\label{subsec:network}
Protocols and topologies are two main factors that affect network performance.
Besides general TCP/IP \textit{protocol}, distributed training often uses RDMA (Remote Direct Memory Access), such as RoCE or Infiniband, for less overhead and higher bandwidth. 

The traffic flow within the network is also closely related to the network \textit{topology}. 
Common network topologies used for distributed training include Fat-tree and its variants, Torus, as well as Ring and Full-Mesh topologies. 
These topologies can be combined according to actual needs. 
For example, the NVLink topology of Nvidia's DGX-1\footnote{https://images.nvidia.com/content/pdf/dgx1-system-architecture-whitepaper1.pdf\label{foot:dgx}} 
is a combination of Ring and Full-mesh. 

\subsection{Three-Layer Paradigm of Communication Optimization}\label{subsec:comm_in_arc}
Parallelization strategy, CCL, and network form a \textit{three-layer paradigm} of communication optimization for distributed training, as shown by the colored rectangles in Fig.~\ref{fig:arc-new}. All of them directly or indirectly affect the communication performance of the training process.
Different \textit{parallelization strategies} determine the primarily used collective communication primitives in task graphs, which affect the traffic pattern. 
Various implementations of a communication primitive in \textit{CCL} directly affect the network traffic. 
The same traffic demand often exhibits different performance under different \textit{network} infrastructures.
The topologies designed for distributed training are closely related to the algorithm implementation of collective communication primitives.
For example, the execution process of the Ring-based All-Reduce algorithm results in a ring communication mode. 
That is, each node transmits data to its logical neighbors while all the nodes in the communicator (a communicator is a set of nodes used to perform a collective communication task) form a ring. 
Nvidia's DGX-1\textsuperscript{\ref{foot:dgx}} and Google's Torus~\cite{jouppi2023tpu} topology both contain numerous ring structures, which are well-suited to satisfy the communication needs of Ring-based All-Reduce. 

In Section~\ref{sec:adv}, we will introduce the recent advances of communication optimization in each of the three layers.
We find that inter-layer collaboration is still limited and is a direction worth exploring.
Therefore, we advocate a new \textit{five-layer paradigm} in Section~\ref{sec:oppo}, which includes inter-layer schedulers to achieve more efficient communication.

\section{Overview of Recent Advances}\label{sec:adv}

\begin{table*}[t]
    \caption{Advances on Communication Optimization in Distributed Deep Neural Network Training}
    \label{tab:advances}
    \renewcommand{\arraystretch}{1.5}
    \resizebox{\textwidth}{!}{
    \begin{tabular}{|m{1.4cm}|m{0.7cm}|m{2.4cm}|m{3.35cm}|m{1.2cm}|m{0.7cm}|m{2.0cm}|m{2.5cm}|}
    \hline
    \multirow{2}{*}{\makecell*[l]{Specific\\works}} & \multirow{2}{*}{\makecell*[l]{Layer}}     & \multirow{2}{*}{Focus}                                                                    & \multirow{2}{*}{\makecell*[l]{How to reduce\\ exposed communication}}                                                                             & \multirow{2}{*}{\makecell*[c]{Time for \\decision}}         & \multirow{2}{*}{\makecell*[l]{Scala-\\bility}}    & \multirow{2}{*}{\makecell*[l]{Beneficial\\effect}}         & \multirow{2}{*}{Deployability}                       \\
                                                    &                                           &                                                                                           &                                                                                                                                                   &                                                             &                                                   &                                                            &                                                      \\ \hline
    Megatron-lm~\cite{shoeybi2019megatron}          & \multirow{8}{*}{\makecell*[l]{Para.}}     & Intra-layer model parallelism                                                             & Reduce traffic by removing a synchronization point                                                                                                & Offline                                                     & High                                              & 74\% linear scaling on 512 GPUs                            & Integrated in PyTorch                                \\ \cline{1-1} \cline{3-8} 
    Alpa-Comm~\cite{zhuang2023optimizing}           &                                           & ``Cross-mesh resharding" problem                                                          & Reduce inter-node traffic and overlap with computation                                                                                            & Offline \& Online                                            & Mid                                               & 1.1x training speedup on GPT                               & 900 lines of C++ and 2.5K lines of python            \\ \cline{1-1} \cline{3-8} 
    PTD-P~\cite{narayanan2021efficient}             &                                           & Combination of different parallelism                                                     & Overlap with computation by a interleaved pipeline                                                                                                & Offline                                                     & High                                              & 502 petaFLOP/s on 3072 GPUs                                & Integrated in PyTorch                                \\ \cline{1-1} \cline{3-8} 
    Lina~\cite{li2023accelerating}                  &                                           & All-to-All bottleneck in MoE parallelism                                                  & Overlap All-Reduce comm. with computation                                                                                                         & Offline                                                     & Mid                                               & Up to 1.73x training speedup                               & Integrated in PyTorch                                \\ \cline{1-1} \cline{3-8} 
    Janus~\cite{liu2023janus}                       &                                           & Data-centric paradigm for MoE parallelism                                                 & Reduce All-to-All traffic and overlap with computation                                                                                            & Offline \& Online                                            & High                                              & Up to 2.4x training speedup                                & Integrated in PyTorch                                \\ \hline
    NCCL                                            & \multirow{9}{*}{\makecell*[l]{CCL}}       & Comm. primitive implementation                                                            & Select efficient communication primitive algorithms                                                                                               & Offline \& Online                                            & High                                              & Widely used infrastructure                                 & Integrated in PyTorch and TensorFlow                 \\ \cline{1-1} \cline{3-8} 
    Blink~\cite{wang2020blink}                      &                                           & \multirow{2}{*}{\makecell*[l]{Comm. primitive \\optimization for \\underlying topology}}  & \multirow{5}{*}{\makecell*[l]{Overlap flows by fine-grained \\scheduling to improve link\\ utilization and shorten \\primitive completion time}}  & Online                                                      & Low                                               & Up to 1.4x training speedup                                & Integrated in PyTorch and TensorFlow                 \\ \cline{1-1} \cline{5-8} 
    SCCL~\cite{cai2021synthesizing}                 &                                           &                                                                                           &                                                                                                                                                   & Offline                                                     & Low                                               & 1.14x to 2.2x training speedup                             & Integrated in MSCCL (Microsoft CCL)                  \\ \cline{1-1} \cline{3-3} \cline{5-8} 
    TACCL~\cite{shah2023taccl}                      &                                           & Synthesizing comm. primitive with human sketch                                            &                                                                                                                                                   & Offline                                                     & Mid                                               & Up to 2.36x training speedup on BERT                       & Integrated in PyTorch                                \\ \cline{1-1} \cline{3-8} 
    SYNDI-CATE~\cite{mahajan2023better}             &                                           & Co-optimization of runtime schedule and execution                                         & Overlap primitives to promote link utilization                                                                                                    & Offline \& Online                                            & Mid                                               & 1.21x to 1.74x training speedup                            & Integrated in PyTorch                                \\ \hline
    TPUv4~\cite{jouppi2023tpu}                      & \multirow{3.5}{*}{\makecell*[l]{Net.}}      & Supercomputer for machine learning                                                        & Efficient topology for distributed machine learning                                                                                               & Offline                                                     & High                                              & Nearly 10x faster overall                                  & Up to 4096 chips connected by OCSs                   \\ \cline{1-1} \cline{3-8} 
    TopoOpt~\cite{wang2023topoopt}                  &                                           & Co-optimization for topology and parallelization strategy                                 & Match topology with parallelization strategy                                                                                                      & Offline                                                     & High                                              & Up to 3.4x training speedup                                & With FlexFlow's training engine and modified NCCL    \\ \hline
    \end{tabular}
    }
\end{table*}

Communication optimization for distributed training encompasses numerous aspects. 
We delve into three of the most pertinent elements: parallelization strategy, CCL, and network. 
In this section, our focus is confined to the contributions of a specific study pertaining to communication optimization, that is, how to reduce \textit{exposed communication} (the communication time that cannot overlap with computation time) in distributed training. 
Table~\ref{tab:advances} provides an overview of the research advances with some primary concerns.
In the table, we use Para., CCL and Net. to represent Parallelization Strategy, Collective Communication Library and Network respectively.
We evaluate the studies from many aspects such as \textit{time for decision}, \textit{scalability}, and \textit{deployability}.
\textit{Time for decision} indicates whether it plays a role during training (online) or before training (offline), which is crucial for practical applications. 
\textit{Scalability} refers to the ability to be applied to large scenarios.
We also show representative performance metrics under specific experimental settings in \textit{beneficial effect}.

\subsection{Parallelization Strategy}
The communication overhead of distributed training stems from the data synchronization requirements brought about by parallelization.
Different parallelization strategies determine different communication types, which in turn bring about different network traffic patterns~\cite{wang2023topoopt}.

The most straightforward and most widely used parallelization strategy is \textit{data parallelism}. 
As mentioned in Section \ref{sec:arc}, multiple copies of the same model are replicated across different nodes in data parallelism. 
Each iteration requires all nodes to synchronize gradients, which are used to update global parameters. 
Consequently, data parallelism results in a significant amount of All-Reduce communication.

However, due to memory constraints, very large models cannot be trained using only data parallelism.
To address this, researchers introduced model parallelism, which includes pipeline and tensor parallelism.
Tensor parallelism divides a layer of a model, requiring intensive communication for synchronizing activations/gradients at each layer during propagation.
To reduce the communication traffic in tensor parallelism, researchers from Nvidia proposed an efficient ``intra-layer parallelism" approach called Megatron-lm~\cite{shoeybi2019megatron}.
It achieved 77\% of linear scaling with a 8.3 billion parameters model on 8 GPUs, and 74\% of linear scaling on 512 GPUs. 

To further divide the model by layers, researchers introduced \textit{pipeline parallelism}.
Pipeline parallelism only requires transferring activations/gradients between adjacent nodes in the logical pipeline, which is a point-to-point communication mode that requires less communication compared to data parallelism and tensor parallelism.
Since communication and computation are difficult to overlap in the original pipeline, researchers introduced an interleaved pipeline scheduling strategy~\cite{narayanan2021efficient} where each device is assigned multiple data chunks at a time to gain the chance of communication-computation overlap.
However, pipeline parallelism, as an ``inter-layer parallelism" approach, is not without its flaws. 
It faces issues such as ``synchronous/asynchronous" parameter updates and pipeline bubble problems, 
which can adversely affect the accuracy of the model and the efficiency of training~\cite{shoeybi2019megatron}.  

Various parallelization strategies are often used in conjunction. 
For example, PTD-P~\cite{narayanan2021efficient} combines tensor parallelism, pipeline parallelism and data parallelism to scale to thousands of GPUs. 
It can train a model with 1 trillion parameters
at 502 petaFLOP/s on 3072 GPUs (per-GPU throughput of 52\% of
theoretical peak).
AlpaComm~\cite{zhuang2023optimizing} identifies the ``cross-mesh resharding" problem that arises when combining tensor and pipeline parallelism.
It proposes a communication optimization method based on broadcasting and an ``overlapping-friendly" pipeline scheduling scheme to accelerate end-to-end training.
It shows a 1.1x speedup compared to their state-of-the-art baseline because of overlapping.

The MoE model is an emerging model structure that is gaining popularity due to its high performance, flexibility and fast training/inference speed.
During the training process, each data sample is assigned to several appropriate experts, which inherently has good compatibility with parallelization.
MoE parallelism can be achieved by simply dispersing the ``experts" on different GPUs.
This introduces All-to-All traffic when data is distributed between different GPUs.
Additionally, non-expert part of the model will also introduce All-Reduce traffic during training.
There are many studies on optimizing MoE parallelism.
Lina~\cite{li2023accelerating} uses a scheduling strategy that prioritizes All-to-All traffic and further splits the All-Reduce tasks to overlap with computation as much as possible, thus reducing exposed communication.
Janus~\cite{liu2023janus} studies MoE parallelism acceleration from a new perspective. 
Since the size of expert models and training datasets varies, Janus innovatively proposes a ``data-centric" model that ``moves experts instead of data". 
This significantly reduces the amount of communication traffic up to 16x under certain conditions (e.g., when the scale of experts is smaller than the scale of data) and achieves up to 2.06× speedup compared with MoE training system at the time.

\subsection{Collective Communication Library} 
Communication tasks of distributed training are often implemented by CCL. 
The current predominant NVIDIA Collective Communication Library (NCCL)\footnote{https://developer.nvidia.com/nccl\label{foot:nccl}} 
dynamically selects established algorithms based on different situations. 
However, NCCL's optimization for specific hardware underscores the need for a more universal and adaptable CCL to speed up distributed training across different hardware and topologies. 
Therefore, many researchers take different approaches to implement communication primitives more efficiently. 

For example, Blink~\cite{wang2020blink} dynamically generates optimal communication primitives by packing spanning trees rather than writing and optimizing them manually. 
It uses integer linear programming (ILP) to reduce the number of trees and improve performance, and also leverages heterogeneous communication channels to fully utilize bandwidth. 
At the time, Blink achieved up to 1.4x speedup compared to NCCL in end-to-end DNN training iterations.
SCCL~\cite{cai2021synthesizing} takes a more unified approach to automatically synthesize high-performance communication primitives for a given topology.
It designs a cost model to evaluate the latency and bandwidth cost of algorithms and then searches for algorithms on the Pareto frontier. 
SCCL is 1.14x to 2.2x faster than NCCL's All-Gather on different sizes of data at the time.
However, generating communication primitives is a Mixed Integer Linear Program (MILP) problem, which is unfortunately NP-hard and can require a long time to solve.

Although Blink and SCCL focus on optimization within small heterogeneous topologies and might not be practical to scale to large-cluster scenarios with tens of thousands of GPUs because of complexity, they lay the groundwork for future CCL optimization works.

\begin{figure}[t]
    \centering
    \includegraphics[width=0.4\textwidth]{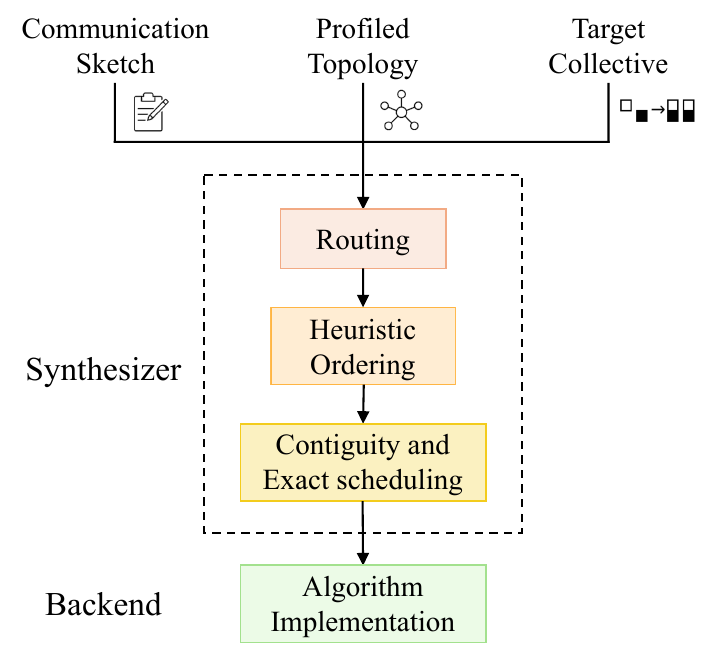}
    \caption{TACCL's workflow. The synthesizer takes as input a communication sketch, profiled topology, and target collective along with synthesizer hyperparameters to generate an algorithm for the collective. The synthesized algorithm is implemented in the hardware cluster using TACCL's backend~\cite{shah2023taccl}.}
    \label{fig:TACCL-arc}
\end{figure}

Based on SCCL, TACCL~\cite{shah2023taccl} introduces a ``human-in-the-loop" approach to reduce the complexity of generating collective communication algorithms. 
The workflow of TACCL is shown in Fig.~\ref{fig:TACCL-arc}. 
TACCL incorporates high-level inputs from an algorithm designer, such as logical topologies, switch hyper-edges, and algorithm symmetry, to efficiently synthesize collective communication algorithms for heterogeneous topologies. 
These human inputs greatly constrain the search space of algorithms, reducing search time to an acceptable amount, in most cases from seconds to a few minutes. 
Compared with NCCL, TACCL achieves up to 2.36x end-to-end training speedup for BERT and 1.94x for Transformer-XL. 

Different from previous methods focused on speeding up single primitives, SYNDICATE~\cite{mahajan2023better} proposes a novel abstraction to overlap primitives in a fine-grained pattern.
It breaks large communication operations into smaller units, thus allowing higher scheduling flexibility. 
SYNDICATE jointly optimizes the implementation and scheduling of communication primitives, enabling parallel execution of primitives.
It improves the speed of training state-of-the-art large models by 21-45\%.
SYNDICATE embodies the idea of cross-layer collaborative design, but is still a preliminary exploration.

\subsection{Network}
Distributed training focuses on GPU-to-GPU communication, which needs to consider intra/inter-host communication protocols. 
Common intra-host communication protocols include the general PCIe protocol and Nvidia's NVLink/NVSwitch,
while inter-host protocols include traditional TCP/IP and more efficient RDMA (e.g., RoCE and Infiniband). 

Besides protocols, the underlying network topology also significantly influences the communication efficiency of distributed training.
For instance, as mentioned in Section~\ref{sec:arc}, Nvidia utilizes a Ring and fully-connected Mesh topology in their DGX series, and Google employs a 3D-Torus topology with optical circuit switches (OCSs) in their TPUv4~\cite{jouppi2023tpu}. 
These topologies contain a large number of ring structures, which enhance the efficiency of Ring-based All-Reduce and Point-to-Point communications (such as pipeline parallelism or broadcast-based hybrid parallelism~\cite{zhuang2023optimizing}). 

However, static topologies can rarely adapt to dynamically changing job requirements, motivating researchers to propose TopoOpt~\cite{wang2023topoopt}. 
TopoOpt takes advantage of the reconfigurability of optical switches, enabling a scheme that co-optimizes the topology and parallelization strategy. 
Nevertheless, current optical switching technology still has issues of high reconfiguration latency, making it challenging to perform dynamic topology adjustments between iterations. 
As such, TopoOpt remains a relatively coarse-grained optimization.
It reconfigures the topology only before the job starts and maintains it until the training is complete.

In conclusion, research on communication optimization for distributed training from a network perspective is still insufficient, which requires further exploration.

\subsection{Collaborative Design in Current Advances}
Current advances involving cross-layer optimization can be categorized into two types. 
The first is passive collaborative design.
For instance, AlpaComm~\cite{zhuang2023optimizing} and Lina~\cite{li2023accelerating} optimize communication for a specific parallelization strategy, necessitating modifications to communication primitives.
Another example is that generative CCL (such as Blink~\cite{wang2020blink}, SCCL~\cite{cai2021synthesizing} and TACCL~\cite{shah2023taccl}) has to be aware of network topologies because their main focuses are to customize collective communication algorithms for specific topologies.
The second type is proactive collaborative design.
For example, SYNDICATE~\cite{mahajan2023better} jointly optimizes the implementation and scheduling of communication primitives, allowing parallel execution of primitives. 
TopoOpt~\cite{wang2023topoopt} collaboratively optimizes network topology and parallelization strategies, obtaining optimized combination strategies through alternating optimization techniques.
However, these works are still preliminary attempts and require further research.

In summary, collaborative communication optimization in distributed training is still a domain that needs further exploration. Section~\ref{sec:oppo} details the importance of collaborative design and potential research opportunities.

\section{Opportunities}\label{sec:oppo}

\begin{figure}
    \centering
    \includegraphics[width=0.95\linewidth]{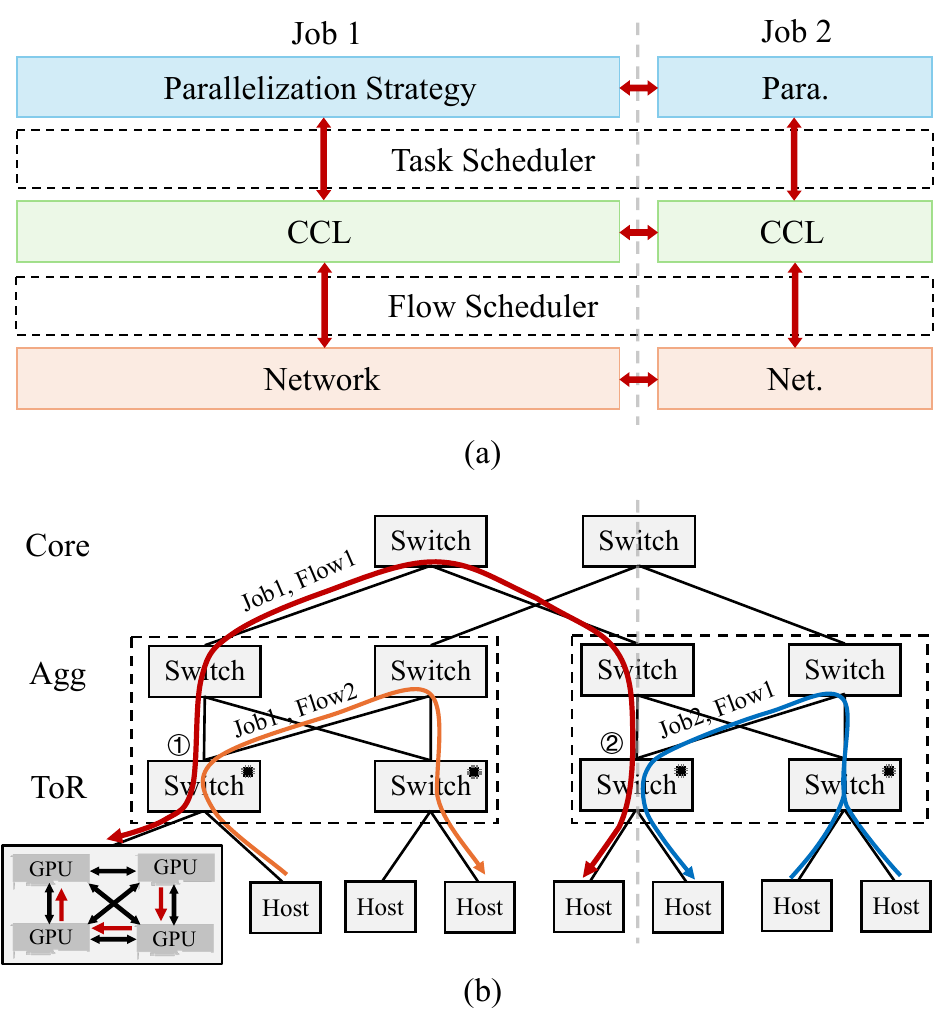}
    \vspace{-10pt}
    \caption{Collaborative design opportunities in distributed training. (a) Communication-efficient five-layer paradigm. (b) Case study of resource competition. The topology is part of a fat-tree with Top-of-Rack (ToR), Aggregation (Agg) and Core three-layer switches. Each host has multiple GPUs and only the left host is detailed. Multiple flows of different training jobs compete for network resources. The switch with the chip logo in the upper right corner represents a programmable switch with in-network aggregation capability.}
    \label{fig:oppo}
\end{figure}

Different from general communication tasks in data centers, the data traffic patterns of distributed training are predictable in most cases~\cite{rajasekaran2024cassini, pan2022efficient}.
Furthermore, the optimization objectives of communication tasks are no longer the general-purpose flow completion time (FCT) but rather the job completion time (JCT).
Therefore, optimizing JCT necessitates considering the dependency between communication and computation tasks and the relationship between different communication tasks instead of just optimizing a single communication task.
These provide motivations and opportunities for collaborative design to optimize efficiency.

In this section, we use a simple case study (Fig.~\ref{fig:oppo}(b)) to illustrate the necessity of collaborative design in distributed training.
We provide research opportunities from two aspects: the communication-efficient five-layer paradigm and the collaboration of heterogeneous resources.

\subsection{Communication-Efficient Five-Layer Paradigm}
In Section~\ref{sec:arc}, we have introduced the current three-layer paradigm of communication optimization for distributed training, which includes Parallelization strategy, CCL, and Network. 
In order to promote cross-layer collaboration, we extend the three-layer communication paradigm to a \textit{communication-efficient five-layer paradigm}, as illustrated in Fig.~\ref{fig:oppo}(a), 
where the red arrows indicate possible information exchange for collaborative optimization.

In the new paradigm, the Parallelization Strategy (Para.), CCL, and Network (Net.) are still the three main layers, while the (logical) schedulers are added as the middleware. 
The task scheduler is to schedule the communication tasks required by the parallelization strategy, and the flow scheduler is to schedule the communication flows generated by CCL.
Standardized interfaces and protocols are required to facilitate seamless communication between layers, which allow lower-layer facilities to perceive the requirements of higher-layer applications and enable higher-layer applications to be aware of the capabilities of lower-layer facilities.
Each layer can use shared information to optimize with a broader perspective, achieving \textbf{``Vertical" co-design}.

For example, \textit{CCL} can generate task-specific communication primitive algorithms based on the global traffic demand from the \textit{parallelization strategy} and the \textit{network} state, thus minimizing JCT.
Furthermore, cross-layer schedulers can be constructed to strategically schedule communication tasks based on traffic load or optimize task graphs according to the traffic patterns of relevant communication primitives. 
In the case study of Fig. ~\ref{fig:oppo}(b), two concurrent flows of Job1 compete for ToR switch at \textcircled{1}, which may lead to congestion. 
However, the ``deadline" of the two flows may be different, and the scheduler can adjust the priority of the flows to avoid blocking the subsequent computation process. 
In other words, the scheduler can consider both the task graphs and the network constraints to optimize the communication tasks, which is an example of ``Vertical" co-design.
This collaborative schedule can minimize JCT for the entire Job1, not only FCT for a certain flow.
Echelon~\cite{pan2022efficient} is a flow scheduling example that embodies ``Vertical" co-design, where novel communication optimization objectives are designed based on the dependency between communication and computation under various parallelization strategies, aiming to optimize communication towards minimizing JCT. 

The schedulers can also perform global scheduling across multiple jobs, achieving \textbf{``Horizontal" co-design}.
In multi-job scenarios, the traffic from numerous concurrent training jobs competes for network resources.
As shown in Fig.~\ref{fig:oppo}(b), flows of Job1 and Job2 compete for the same ToR switch at \textcircled{2}.
Consequently, optimization strategies tailored to a single training job may suffer varying degrees of inefficacy. 
Therefore, it is imperative to employ horizontal co-design at different granularities to fully utilize network resources, ultimately reducing JCT of training tasks. 
A potential solution is to implement dynamic resource manager that allocates resources based on real-time demands.
CASSINI~\cite{rajasekaran2024cassini} attempts horizontal co-design through the perspective of multi-task scheduling. 
It performs ``staggering peak'' scheduling for different jobs, 
which improves the utilization of network resources.
In addition, horizontal collaboration also enables multi-job Service Level Agreement (SLA) management.

\subsection{Collaborate Design for heterogeneous resources}
Considering the heterogeneous network topologies of distributed training and the development of emerging technologies such as in-network aggregation, we provide a prospect for ``Intra-Inter" and ``Host-Net" co-design.

\textbf{``Intra-Inter" co-design} refers to the coordinated utilization of multidimensional network resources, primarily including the ``intra-host" and ``inter-host" network.  
GPUs are typically interconnected via fast PCIe or NVLink within a host, while inter-host connections are established through relatively slower RoCE or InfiniBand.
In this context, the issue of heterogeneous link bandwidths within and between hosts is inevitably confronted.
Let us suppose that Flow1 of Job1 is to implement a Ring All-Reduce among GPUs across two hosts in Fig.~\ref{fig:oppo}(b).
Then the bottleneck of this communication task will be the bandwidth of the inter-host link, which is slower than the intra-host link.
Solving this problem requires real-time sensing and scheduling of heterogeneous link resources.
TACCL~\cite{shah2023taccl} utilizes a profiling approach to perceive the characteristics of heterogeneous links in order to optimize the communication primitive algorithms, which embodies the concept of ``Intra-Inter" co-design.

\textbf{``Host-Net" co-design} primarily refers to the collaboration of computing resources between the end-host and programmable switches within a network. 
In recent years, with the development of programmable switch technology, in-network aggregation has gradually become a new computing paradigm.  
In-network aggregation uses the computing capabilities of programmable switches within the network to reduce the traffic running in the network.
As shown by the blue line in Fig.~\ref{fig:oppo}(b), the two flows of Job2 are aggregated into one flow at the ToR switch, thus reducing further communication overhead.
However, since existing protocols are designed for end-to-end communication, implementing in-network aggregation usually requires the support of new protocols. 
For instance, ATP~\cite{lao2021atp} is an in-network aggregation transmission protocol designed for multi-tenant scenarios, which can make full use of the 
computing capabilities of multi-level programmable switches and degrade to host aggregation when the in-network computing resources are exhausted. 
This approach also reduces the network traffic for a collective communication task.
ATP embodies the concept of ``Host-Net" co-design. 
Nevertheless, in-network aggregation has not been widely applied in the production environment due to its complexity. 
``Host-Net" co-design remains a direction worthy of long-term research.

\subsection{Open Issues}

\noindent $\bullet$ \textbf{Cross-layer information exchange mechanism:} The lack of effective information exchange mechanism across different layers makes collaborative optimization difficult. How to design good protocols to achieve efficient information exchange between different layers is a problem worth studying.

\noindent $\bullet$  \textbf{Utilization of heterogeneous network resources:} The full utilization of heterogeneous network resources is a long-term complex problem, involving the awareness of the real-time status of different network resources and timely scheduling. 
This poses high requirements on the complexity and accuracy of the scheduler.

\noindent $\bullet$  \textbf{Dynamic optimization of communication libraries:}
Current communication optimization strategies are usually offline, and existing generative CCLs are difficult to meet the real-time requirements of online operation. However, network status is dynamic, and communication patterns also change in emerging model structures such as MoE. Therefore, the dynamic optimization of communication libraries is an important research direction.

\noindent $\bullet$  \textbf{Network topology specific for distributed training:}
Current network topologies are usually optimized for general tasks (e.g., fat-tree) or common communication primitive implementations (e.g., Ring-based All-Reduce). 
How to design or generate an efficient network topology for distributed training tasks and introduce dynamic adjustments to the topology according to the change in traffic load is a research direction worth exploring. 

\noindent $\bullet$  \textbf{Communication Optimization with emerging techniques:} With the emergence of new hardware (such as programmable switches, optical switches, etc.), it is important to study the impact of these new hardware on communication optimization and explore how to make full use of these hardware to improve communication efficiency.
\section{Conclusion}
Due to the dramatic increase in the number of parameters of large-scale deep neural network models, it is imperative to develop effective communication optimization methods in distributed training. 
In this article, we first introduce a general distributed training architecture from the communication perspective and summarize the current three-layer paradigm of communication optimization, including parallelization strategy, CCL, and network.
We review the latest advances in the current three-layer paradigm and summarize their important features in a table.
Then, we look forward to future research directions from the perspective of collaborative design. 
We advocate a communication-efficient five-layer paradigm and collaborative design for heterogeneous resources.
Finally, we propose some open issues.
We believe that this article can shed some light on future research on communication optimization for distributed training.
\section*{Acknowledgments}
This work was supported by NSFC Project under Grant 62132009 and Grant 62221003.

\bibliographystyle{IEEEtran}
\bibliography{main}

\vspace{-30pt}
\begin{IEEEbiographynophoto}{Yunze Wei (wyz23@mails.tsinghua.edu.cn)} is currently pursuing a Ph.D. degree with the Department of Computer Science and Technology, Tsinghua University, Beijing, China. He received a B.E. degree in the School of Computer Science and Technology, Huazhong University of Science and Technology, Wuhan, China, in 2023. His current research interests include data center networks and distributed machine learning. 
\end{IEEEbiographynophoto} 

\vspace{-30pt}
\begin{IEEEbiographynophoto}{Tianshuo Hu (huts22@mails.tsinghua.edu.cn)} is currently pursuing a B.E. degree with the Department of Computer Science and Technology, Tsinghua University, Beijing, China. His current research interests include distributed machine learning and data center networks.
\end{IEEEbiographynophoto}

\vspace{-30pt}
\begin{IEEEbiographynophoto}{Cong Liang (liangc23@mails.tsinghua.edu.cn)} received the B.E. and M.E. degrees from Tsinghua University, Beijing, China, in 2020 and 2023. He is currently working toward his Ph.D. degree in the Department of Computer Science and Technology at Tsinghua University. His current research interests include data center networks and network optimization for distributed machine learning.
\end{IEEEbiographynophoto}

\vspace{-30pt}
\begin{IEEEbiographynophoto}{Yong Cui (Member, IEEE) (cuiyong@tsinghua.edu.cn)}
  received the B.E. and Ph.D.
  degrees in computer science and engineering
  from Tsinghua University, China, in
  1999 and 2004, respectively. He is currently
  a Full Professor with the Computer
  Science Department, Tsinghua University.
  His major research interests include mobile
  cloud computing and network architecture.
  He served or serves on the editorial boards of the IEEE
  TPDS, IEEE TCC, IEEE Internet Computing, and the IEEE Network.
\end{IEEEbiographynophoto}

\end{document}